\documentclass[journal]{IEEEtran}

\usepackage{colortbl}
\usepackage[pdftex]{graphicx}

\ifCLASSINFOpdf
 
\else
 
\fi

\usepackage[cmex10]{amsmath}

\usepackage{amsmath, amsthm, amssymb, amsfonts, array}
\usepackage{graphicx}
\usepackage{caption}
\usepackage{subcaption}

\begin{document}
\title{Highly Dynamic Spectrum Management within Licensed Shared Access Regulatory Framework}
%
%
%

\author{Aleksei~Ponomarenko-Timofeev, 
				Alexander~Pyattaev, 
				Sergey~Andreev$^{\dagger}$, 
        Yevgeni~Koucheryavy,\\ 
        Markus~Mueck, 
        and~Ingolf~Karls
\thanks{A.~Ponomarenko-Timofeev, A.~Pyattaev, S.~Andreev, and Y.~Koucheryavy are with the Department of Electronics and Communications Engineering, 
Tampere University of Technology, FI-33720 Tampere, Finland.}
\thanks{M.~Mueck and I.~Karls are with Intel Mobile Communications, Germany.}
\thanks{The work of the third author is supported with a Postdoctoral Researcher grant by the Academy of Finland as well as with a Jorma Ollila grant by Nokia Foundation.}
\thanks{$^{\dagger}$S.~Andreev is the contact author: Room TG417, Korkeakoulunkatu 1, 33720, Tampere, Finland (+358 44 329 4200); e-mail: sergey.andreev@tut.fi}
\thanks{\textbf{Open Call}; COMMAG-15-00508, Guest Editor: Zoran Zvonar}
}

%

\maketitle

\begin{abstract}
Historical fragmentation in spectrum access models accentuates the need for novel concepts that allow for efficient sharing of already available but underutilized spectrum. The emerging Licensed Shared Access (LSA) regulatory framework is expected to enable more advanced spectrum sharing between a limited number of users while guaranteeing their much needed interference protection. However, the ultimate benefits of LSA may in practice be constrained by space-time availability of the LSA bands. Hence, more dynamic LSA spectrum management is required to leverage such real-time variability and sustain reliability when e.g., the original spectrum user suddenly revokes the previously granted frequency bands as they are required again. In this article, we maintain the vision of highly dynamic LSA architecture and rigorously study its future potential: from reviewing market opportunities and discussing available technology implementations to conducting performance evaluation of LSA dynamics and outlining the standardization landscape. Our investigations are based on a comprehensive system-level evaluation framework, which has been specifically designed to assess highly dynamic LSA deployments.
\end{abstract}

\section{Introduction and background}

\subsection{Current spectrum access and management models}

Over the years, radio spectrum has become a critical resource for numerous purposes: from economic and social to cultural and scientific. However, its management has largely remained unchanged in the course of the past three decades due to the underlying complexity of the process and insufficient maturity of radio technology. Along these lines, various distinct approaches to spectrum management have historically taken shape.

\subsubsection{"Command-and-control" spectrum management} This age-old antiquated paradigm executes static spectrum allocation. Accordingly, a regulatory body assigns a frequency band to a particular entity while imposing strict constraints on such use. Naturally, this approach led to barriers in spectrum access bringing along difficulties to meet the increasing demand for wireless spectrum based services. In addition, the corresponding assignment of frequency bands never relied on market mechanisms, hence resulting in very low economic profits. Conventionally, spectrum ownership rights have been granted as the result of so-called "beauty contests" and required considerable lobbying to regulation authorities.

\subsubsection{Exclusive use of spectrum} This model is centered around a long-term ($15$ to $30$ years) spectrum band \textit{license} awarded to utilize a particular band. Correspondingly, the resulting use is subject to certain well-defined rules, such as maximum power levels and geographical coverage. Exclusive licenses empower their respective owners (e.g., cellular network operators) with unrestricted interference management capabilities thus enabling quality-of-service (QoS) guarantees, but at the same time impose high market entry barriers (i.e., billions of Euros). As opposed to the legacy "beauty contests", assignment was transformed by sales of spectrum: most regulators have now adopted market-centric approaches (e.g., auctions) to redistribute frequency allocations.

\subsubsection{Shared use of primary licensed spectrum} In this concept, the frequency bands of a licensed owner (named primary user) are shared by a non-license holder (named secondary user). Importantly, access by secondary user may sometimes occur without notifying the primary user and requires the respective protection of the latter, such that the intended operation of primary communication is not deteriorated. In this regard, there has been a recent surge in software-defined radio technologies, cognitive and adaptive radio networks, as well as reconfigurable networking to enable the intended dynamic spectrum access (e.g., in TV white space). However, the fundamental limitation of this form of access is in that it is unclear how the secondary user may deliver reliable QoS guarantees over such shared spectrum.

\subsubsection{Shared use of unlicensed spectrum} When a spectrum band is allowed for "open access", no entity can claim its exclusive use and the target spectrum should be made fairly accessible to everyone. The example of such spectrum usage is industrial, scientific, and medical (ISM) bands, where multiple potential users (e.g., medical and sensor devices, microwave ovens, cordless phones, WiFi networks, etc.) may access the spectrum without external regulation. While such unregulated access significantly lowers market entry barriers, it also produces uncontrolled wireless interference and, consequently, makes it extremely hard to meet the desired QoS guarantees. In addition, a multitude of spectrum sharers may lead to a situation when none of the users achieve their expected benefit. This is a very likely course of development today, given the increasing popularity of WiFi and the corresponding emphasis of network operators on different forms of WiFi offloading.

\subsection{Transformation of global wireless landscape}

To overcome the long-standing effects of fragmentation in spectrum access models, there is a pressing demand for novel frameworks allowing for efficient sharing of already available but underutilized spectrum. The need for this change is becoming increasingly urgent as the pressure on the radio spectrum is steadily building, largely due to the unprecedented explosion in wireless traffic. Indeed, recent forecasts by Cisco predict the growth in mobile data demand at a rate of nearly 60\% over the following $5$ years, which brings along the $10$-fold overall increase. In this regard, the past traffic growth predictions look overly optimistic in that they heavily underestimate the mobile data acceleration~\cite{Mue14}.

As data from mobile and wireless devices is expected to soon exceed traffic from wired equipment, the 4G networks of today face the risk of "capacity crunch". To this end, the forthcoming 5G technologies offer a range of decisive improvements in cell capacity~\cite{And14}. However, more efficient use of existing spectrum will not solely be sufficient to achieve the needed factors of 1000 to 10,000-fold improvement. These targets by the year 2020 are impossible without the availability of additional frequency resources, which will be required for a range of spectrum-hungry technologies: from conventional mobile-to-infrastructure links to complementary device-to-device and multi-hop communication, as well as wireless front- and backhauling.

Regrettably, given that the traditional approach of re-purposing spectrum is reaching its limits (especially on bands below 6 GHz), it is unlikely that more contiguous and broader microwave frequencies will be made available any time soon. At the same time, whereas radio spectrum may be saturated during peak hours and/or in crowded locations, there is presently an extreme variability of load across time and space. Hence, this dynamics may be exploited to manage spectrum more efficiently, especially given the fact that the average traffic grows at a much slower rate than the busy-hour traffic. Currently, though, there are no feasible options to manage spectrum on such small-scale spatio-temporal granularity, which calls for new approaches to spectrum policy and allocation methods.

\begin{figure}[!ht]
\centering
\includegraphics[width=0.7\columnwidth]{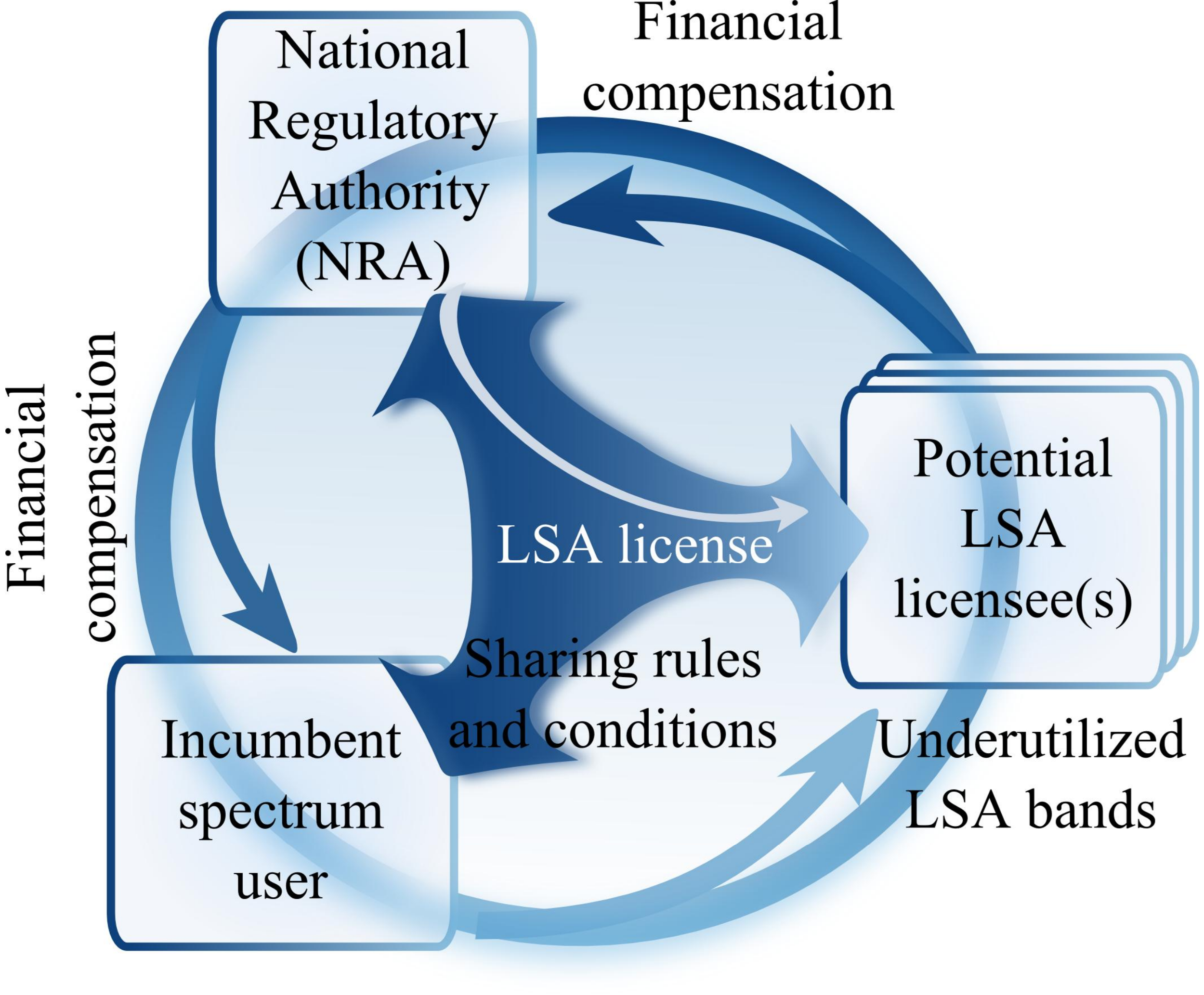}
\caption{LSA regulatory framework: key stakeholders}
\label{fig:interactions}
\end{figure}

In light of the above, it appears that the shared use of spectrum becomes unavoidable even for those who have conventionally enjoyed exclusive access rights~\cite{Cha15}. However, the existing forms of spectrum sharing (in primary licensed or unlicensed spectrum, see above) do not offer the much needed interference protection, thus resulting in insufficient reliability, QoS guarantees, and predictability of operation. By contrast, the emerging Licensed Shared Access (LSA) regulatory concept (see Fig.~\ref{fig:interactions}) allows for more advanced spectrum sharing between a limited number of entities with carefully-defined usage rights -- combining the benefits of "command-and-control" spectrum management with a flexible and innovative market-friendly approach. 

Broadly, LSA enables \textit{authorized} spectrum sharing by allowing at least two users, named the \textit{incumbent} (i.e., the current holder of spectrum rights) and the LSA \textit{licensee} (i.e., the temporary user of spectrum) respectively, to access the same frequency bands in a licensed pre-determined manner following a well-defined mutual agreement~\cite{Buc14}. In other words, LSA guarantees that the incumbent retains spectrum access rights anytime, anywhere and that the LSA licensee(s) will refrain from using this spectrum when in need by the incumbent (or at least will not disrupt the incumbent's operation).

Under the LSA's "individual licensing regime", sharing agreements need to guarantee high predictability in terms of spectrum access for all the involved parties:

\begin{enumerate}
	\item For the incumbent(s), LSA leverages additional economic benefits from underutilized spectrum without imposing any significant operational restrictions on its expected use. 
	\item For the national regulator, LSA harmonizes spectrum usage opening path to its optimization via controlled sharing as an alternative to permanent segmentation.
	\item For the licensee(s), LSA delivers additional frequencies at more affordable costs together with predictable QoS guarantees due to coordinated interference.
\end{enumerate}

However, the licensee's benefits from LSA may in practice be constrained by space and time availability of the LSA bands. As long as LSA usage remains static, it should suffice that a dedicated exclusion zone or time is created to protect the incumbent's use of spectrum. On the other hand, in case of dynamic geographic/temporal LSA sharing, on-demand authorization of the LSA licensee(s) is required as a consequence of real-time restrictions imposed by the incumbent. While such dynamic LSA systems are more complex to build and maintain, they also unlock higher potential performance benefits. In what follows, we concentrate on highly dynamic LSA operation allowing licensee's spectrum access over a particular frequency, time, and location. To this end, we offer our vision of the required functionality for the LSA architecture to support such dynamics. In addition, we summarize our recently-completed system-level study of LSA performance with a dedicated set of tools that we contribute to make conclusions across a wide range of LSA-centric use cases and scenarios.


\section{LSA system architecture and implementation}

\subsection{Use cases and market opportunities}

To ensure pragmatic and efficient LSA operation providing the desired spectrum access flexibility and harmonization, it is crucial to identify viable use cases and scenarios of its application. 

\subsubsection{Mature operator markets} First and foremost, already today LSA may benefit mobile network operators (MNOs) with mature 3G markets, but lacking 4G coverage and capacity benefits due to lengthy spectrum refarming process. In such markets, where players are typically reluctant to alter their existing business strategies, LSA may change the rules of the competition by allowing smaller MNOs to quickly augment their capacity and coverage. 

\subsubsection{Smaller and virtual operators} Going further, the larger dominating MNOs owning exclusive spectrum licenses may be challenged by smaller MNOs, which had in the past restricted business opportunities due to very little exclusive spectrum. However, dominating MNOs can also strengthen their market positions by acquiring extra LSA bands. In addition, non-MNO players, such as \textit{virtual} network operators, may proliferate on the market, thus reshaping the existing business ecosystem.

\subsubsection{Mobile broadband services} LSA may also support the forward-looking governmental plans to increase adoption of public services over mobile broadband. Indeed, as predicted by many sources, the novel types and higher numbers of wireless services are very likely to be mushrooming at around 2020. Supported by LSA, the emerging 5G trends may include mobile ultra high-definition holography and multimedia-based immersion, large-scale augmented and virtual reality, big data processing, as well as public safety and disaster relief.

\subsubsection{Rural and machine-type markets} Furthermore, LSA holds significant promise for markets with large rural population, as well as for machine-to-machine, wearable, and Internet-of-Things (IoT) markets. To this end, LSA may help leverage the available secondary spectrum in the areas with low population densities. 

\subsubsection{Ultra-dense heterogeneous networks} Finally, LSA also has the potential to aid the deployment of ultra-dense networks based on multi-radio small cells. Overall, the latest analysis of frequency requirements by ITU Radiocommunication Sector (ITU-R) indicates a significant bandwidth demand across today's heterogeneous network deployments. Consequently, over 1000 MHz of new spectrum is currently required, and more efficient spectrum utilization frameworks, such as LSA, are an important building block to enable certain well-defined scenarios.

While a separate LSA business case may be difficult to identify, it can be foreseen that LSA will become one of the potential dynamic spectrum access modes together with exclusive access, co-primary shared access, authorized shared access, unlicensed access, and, perhaps, other options in future 5G systems. Hence, as exclusive access will continue to remain the preferred method of spectrum usage by 5G-grade MNOs, we believe that LSA will be increasingly employed as a complementary approach in conjunction with other spectrum access alternatives, such as unlicensed WiFi in 2.x and 5.x GHz bands, TV bands below 800 MHz, unlicensed cellular access in 1800 MHz, etc.

Naturally, LSA principles are based on voluntariness, where the regulator is not expected to force incumbent(s) to accept sharing. Instead, driven by their economic benefits the incumbent(s) are expected to provide the LSA licensee(s) with access to a part of their spectrum at certain locations and times. In addition, rules must be defined allowing the incumbent to \textit{revoke} such granted spectrum should it be required again (or if the licensee is causing harmful interference to the incumbent), and the respective mechanisms are considered in what follows.

\subsection{Prospective LSA system architecture}

\begin{figure*}[!ht]
\centering
\includegraphics[width=1.9\columnwidth]{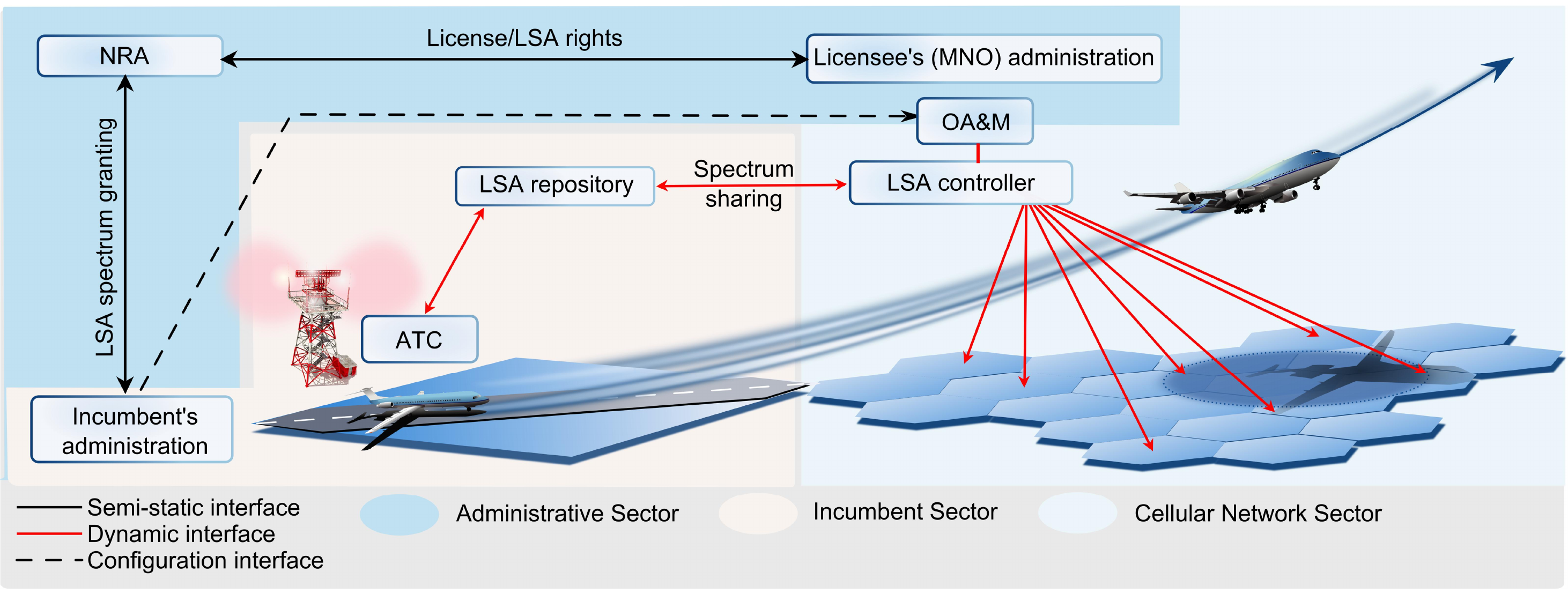}
\caption{Envisioned LSA architecture and motivating scenario}
\label{fig:scenario}
\end{figure*}

As follows from the above, the envisioned LSA ecosystem assumes an intricate interplay between the national regulatory authority (NRA), the incumbent(s), including both governmental and commercial entities, and the potential LSA licensee(s). To define a simple and easy-to-deploy sharing framework, as well as determine appropriate rights of its use, these stakeholders need to engage into intensive bi- and tri-lateral dialogs~\cite{Aho14}. This should allow cellular operators to leverage additional spectrum on a secondary basis, with exclusive and guaranteed access over certain time, frequency, and geographical area. To this end, the prospective LSA system architecture~\cite{TS235} features the LSA repository, the LSA controller, and the mobile wireless communication network operations, administration, and management (OA\&M) entity (see Fig.~\ref{fig:scenario}).

\subsubsection{LSA repository} is essentially a database that may include various information on both the incumbent and the licensee(s). In particular, it needs to store the up-to-date space, time, and frequency information on incumbent's spectrum utilization. Accordingly, the repository is primarily responsible for delivering information on spectrum availability and associated conditions, but may also add safety margins and even deliberate distortions to such data -- the incumbent may not be willing to disclose precise information due to its sensitive nature. The management of the LSA repository may be performed by the NRA or the incumbent directly, or can be delegated to a trusted third party. 

\subsubsection{LSA controller} generally manages access to the spectrum made available to the LSA licensee based on sharing rules and information on the incumbent's use provided by the LSA repository. There is typically a direct link between the LSA controller and at least one LSA repository, which allows for secure and reliable information transfer and requires a standardized communication interface. Correspondingly, after the incumbent's spectrum use information from the repository has been combined with the sharing rules built upon the current LSA usage rights, the controller evaluates the LSA spectrum availability and provides the respective grant to the LSA licensee.

\subsubsection{OA\&M entity of the LSA licensee's mobile network} performs the actual management of the LSA spectrum by issuing the radio resource management (RRM) commands based on the information received from the LSA controller. These RRM commands, after they have been delivered to the MNO's base stations, enable user equipment (UE) to either transmit on the LSA spectrum or hand over to another frequency band subject to LSA spectrum availability, QoS requirements, or data plan preferences. In addition, OA\&M can help the associated base stations with channel and/or transmit power level selection. 

The considered LSA system design enables efficient transition from relatively static to significantly more dynamic LSA operation. Indeed, for static incumbents bound to a particular location and time (e.g., a military base or a TV studio) the resulting interference could be controlled by simple pre-planned exclusion methods. However, in case of a dynamic incumbent (e.g., a radar system or a broadcasting service provider), a significantly more capable low-latency interface between the LSA repository and the LSA controller should become available. It needs to allow for near real-time coordination between the incumbent(s) and the LSA licensee(s), as well as for timely revocation of the LSA frequency bands by the incumbent in case of emergency or excessive interference from the licensee. We thus continue with reviewing the available implementation options for the dynamic LSA system.

\subsection{Technology aspects and LSA implementation design} 

In the first place, LSA needs simple mechanisms allowing the users of a licensee MNO to efficiently enter and vacate the LSA spectrum. For example, after radio access network (RAN) begins advertising the availability of the LSA band, the idle-mode UE may follow the standard re-selection procedures to move to the LSA frequencies. However, such decision is user-centric in nature -- it may cause lengthy delays and uncertainty in intended LSA operation and hence may not be preferred by the MNOs. An alternative network-centric solution is to directly handover the connected-mode UE to a certain component carrier within the LSA frequencies. Unfortunately, in presently deployed cellular networks, such as 3GPP LTE Release 8 and 9, the UE may only use one component carrier at a time, which naturally limits the possibility to employ both primary licensed band and LSA band for increased reliability. 

Starting with its Release 10, LTE technology defines carrier aggregation (CA) mechanism that essentially enables the utilization of several component carriers simultaneously. Given that CA also remains under the full control of the network, it is more efficient and robust since the UE does not in fact change its underlying operating band. More importantly, CA provides means to implement LSA already today, without significant modification of existing MNO deployments. The downside of CA is, however, the need for higher signaling overhead and research is currently under way to improve CA operation for LSA. In addition, it is expected that practical LSA deployments would require a range of new dedicated mechanisms taking into account the radio technology used for incumbent's transmissions, such as RRM, interference mitigation, load balancing, and traffic steering schemes, together with respective network planning modifications.

Along these lines, a crucial underlying LSA mechanism is the possibility of the incumbent to revoke the spectrum band while the LSA license is still effective, which may be required for the reasons discussed previously. To do so, the incumbent needs to inform the LSA repository of the change in its spectrum availability by sending what is known as "evacuation request" via a dedicated interface. Importantly, to enable LSA spectrum allocation/revocation in dynamic on-demand fashion, the LSA controller needs to have a direct low-latency and high-reliability interface with the corresponding control entity in the MNO's core network, which is, in turn, connected to e.g., the serving gateway or the mobile management entity via the S1 interface. This should allow for a more dynamic response to any changes in licensing and offer better predictability in licensee's spectrum usage.

In what follows, we study a characteristic LSA use case, where the incumbent that owns a spectrum license over a large geographical area requires its frequency resource only occasionally, for small and localized portions. We also assume a reasonable cellular network presence in the same area, and the respective MNO has established a direct high-speed interface that enables the incumbent to constrain the interference generated to it by the cellular network explicitly, without additional lengthy negotiations.

In our example scenario, an airport leases its telemetry spectrum to a mobile network (see Figure~\ref{fig:scenario}), which uses this spectrum exclusively, until an airplane needs to be tracked by the air traffic control (ATC). When it happens, the ATC instructs the MNO to restrict its interference around the position of the airplane, as to allow the telemetry collection. This is feasible since the location of the airplane is known by the ATC. In the end, it is up to MNO to decide how to implement the imposed interference constraints, giving the MNO an opportunity to smoothly transition its UEs from the LSA band to the primary licensed band whenever necessary. The scale of such interference management may range from small transmit power adjustments to full "shutdown" of the LSA bands, and has a number of associated challenges: 
 

\begin{itemize}
	\item An exact location of the recall source (e.g., an airplane) should be known, as well as the corresponding radio propagation model to guarantee efficient isolation.
	\item The control interface has to operate with adequate dynamics to keep up with fast-moving objects (such as high-speed trains and airplanes), as to avoid excessive reservations.
	\item The control interface must be sufficiently reliable, as to not affect the operational reliability of the incumbent(s).
\end{itemize}

In what follows, we construct a realistic LSA scenario to exemplify the operation of such highly dynamic system, and provide numerical insights into its expected performance. Note that this usage model, while does not intend to highlight all of the LSA features, is representative and may be adopted today on vast geographical areas, thus constituting a viable business case.


\section{Modeling highly dynamic LSA operation}

\subsection{Characteristic LSA scenario}

Our motivating scenario for the dynamic LSA operation is demonstrated in Fig.~\ref{fig:scenario}. We first note that the majority of today's airports have rather small airfields (see, e.g., \verb+http://www.transtats.bts.gov/airports.asp+) and may not even have a tower. For such small airports, it could be relatively expensive to have dedicated radio resources for the ATC functioning -- these have to be controlled carefully in order to operate. Further, it is not sufficient to only reserve spectrum resources around the airport premises; they must also be available in larger surrounding areas, as long as the airplanes remain relatively close to the ground. Indeed, today's airplanes require time and space to take off and land, hence the resulting exclusion zones end up being vast, up to $25$ km in radius.

In our practical small airport scenario, the airplanes do not arrive/depart every minute. Realistically, we expect an airplane every $10-20$ minutes, or sometimes even less often than that. Moreover, since there are not too many airplanes in the air, they cannot receive signal/interference from the entire exclusion zone, only from a smaller part of it. As a result, the majority of the exclusion zone could often be underutilized, and the corresponding spectrum may thus be available to share with a MNO. Naturally, the cellular network/users would then need to adjust the transmit power based on where the airplanes are in real-time, so as to guarantee the required radio channel quality for the ATC operation. We are primarily interested in the detailed performance analysis of such a system revealing the degrees of the adequate interference control measures by the operator.

To facilitate the corresponding evaluation, several clarifying assumptions have to be adopted. A single runway is focused on, with airplane arrivals and departures separated by at least $5$-minute time intervals. As a consequence, there are never two airplanes in the exclusion zone in our model. Further, all of the airplanes follow the same ascent profile, and only employ telemetry at lower altitudes. The frequency bands in use by the telemetry are shared between the ATC system and the cellular system with LSA. Given that the telemetry transmission is bidirectional, the airplane receiver must be protected from the interference produced by the licensee MNO. 

To this end, the co-located cellular network operates in cooperation with the ATC and attempts to utilize the shared band in the exclusion zone whenever this does not cause excessive interference to the ATC system. Importantly, the base stations (BSs) have directional antennas with downtilt, providing at least $20$ dB isolation between their radiation and the airplane in the air. Hence, we investigate the more interesting case when the LSA band is employed for uplink UE communication to augment the existing primary licensed band in use by the operator. Based on the above considerations, we construct an evaluation scenario in what follows.

\subsection{Proposed system-level evaluation methodology}

Our below evaluation concentrates on the transient period when the airplane e.g., takes off from the runway and travels through the cellular network in the immediate vicinity of the airport. A similar, but reverse pattern would be observed during landing, and we do not evaluate such situation here to avoid redundancy.

\begin{table}[!ht]
\centering
\caption{Simulation scenario parameters}
{\scriptsize
\begin{tabular}{|c|c|}
\hline 
Description & Value\tabularnewline
\hline 
\hline 
\multicolumn{2}{|c|}{Airplane parameters}\tabularnewline
\hline 
Airplane takeoff speed & 65 m/s\tabularnewline
\hline 
Interference threshold ($I_{0}$) & -85 dBm over 10 MHz\tabularnewline
\hline 
Airplane ascent/glide slope & 7 deg\tabularnewline
\hline 
Airplane acceleration & 5 m/s$^{2}$\tabularnewline
\hline 
Air-ground propagation model & Free space\tabularnewline
\hline 
Observation period & 60 s\tabularnewline
\hline 
\multicolumn{2}{|c|}{Cellular parameters}\tabularnewline
\hline 
Cell radius ($R$) & 288 m\tabularnewline
\hline 
Operator's licensed band radio network plan & 1x3x3\tabularnewline
\hline 
LSA band radio network plan & 1x1x1\tabularnewline
\hline 
Cellular scheduling policy & Proportional-Fair\tabularnewline
\hline 
LTE power control parameters (for licensed) & $\alpha = 1$, $SINR_{tgt}$ = 20 dB\tabularnewline
\hline 
LTE power control parameters (for LSA) & $\alpha = 1$, $SINR_{tgt}$ = 5 dB\tabularnewline
\hline 
Maximum BS transmit power & 35 dBm (directional)\tabularnewline
\hline 
Antenna leakage & -35 dB\tabularnewline
\hline 
Propagation model & ITU Urban Micro\tabularnewline
\hline 
Carrier frequency & 2.1 GHz\tabularnewline
\hline 
BS antenna height & 15 m\tabularnewline
\hline 
BS antenna sidelobe isolation & 20 dB\tabularnewline
\hline 
Shadow fading standard isolation & 3 dB\tabularnewline
\hline 
LSA protective margin ($K$) & 10 dB\tabularnewline
\hline 
\multicolumn{2}{|c|}{UE Parameters}\tabularnewline
\hline 
Traffic pattern & Full-buffer (saturation)\tabularnewline
\hline 
Maximum UE transmit power & 23 dBm (isotropic radiator)\tabularnewline
\hline 
Antenna height & 1.5 m\tabularnewline
\hline 
\end{tabular}}
\vspace{-0.5cm}
\end{table}

In more detail, our characteristic LSA scenario operates as follows. A grid of MNO cells is laid out next to a simulated airstrip with the size of $5 \times 5$ cells. Following the respective 3GPP recommendations, cellular users are uniformly located in the grid with the average density of $10$ UEs per cell. Further, at time $t=0$ the airplane is launched from the airstrip and simulation runs until the airplane leaves the network reaching its cruising altitude. Meanwhile, the UEs transmit their saturated data in uplink causing interference on the airplanes. As discussed previously, cellular BSs are assumed to have near-perfect isolation that prevents them from interfering with the airplane systems in downlink.

Our system-level evaluation is performed for three alternative policies of operation:

\begin{enumerate}
	\item IGNORE policy: the airplane travels through the network receiving all possible interference from it. This is a benchmark policy and corresponds to what would happen if no coordination is to be introduced between the LSA incumbent and the licensee.
	\item SHUTDOWN policy: all the BSs whose UEs have a chance to cause interference on LSA bands are "powered off" (in practice this may correspond to a variety of measures to disassociate users and stop transmission). This solution may seem to be the most straightforward, but has unexpected side-effects as we show below.
	\item LIMIT POWER policy: all the BSs are forced to reduce the corresponding UE's uplink power whenever instructed by the ATC, as to meet the interference constraints. This provides a more flexible and efficient control solution, based on a heuristic approach to only limit transmit power when necessary.
\end{enumerate}

Similar policies of operation may be defined for cellular downlink, except that it would be the BS transmit power that is controlled, not the UE power. For policies $2$ and $3$, the cellular network controller has to first learn which cells should be adjusted in response to the airplane presence. Since the exact details of the propagation environment between the cellular entities and the airplane are not known precisely, we could assume the worst case, which corresponds to free space propagation between any transmitter on the ground and the airplane. This may, in fact, be an accurate model since the atmospheric absorption at 2-3 GHz is minimal, and one can observe line-of-sight communication to the airplane from nearly anywhere on the ground, especially around airports.

%

For the LIMIT POWER policy, we attempt to directly bring the interference below of what is required by the ATC. In our evaluations, we assumed a reasonable respective limit of $I_0$ to be $-90$ dBm/$10$ MHz, which is just over the realistic noise floor. In addition, since we have more than a single transmitter interfering with the airplane, we need to introduce a protective margin. For this scenario, as our analysis shows, a sufficient margin is $K=10$ dB, so that all the interference estimates are increased by $K$. For the SHUTDOWN policy, we need to simply "turn off" any BS and its associated UEs as commanded by the ATC. For the IGNORE policy, nothing extra related to power control needs to be done.

The constructed simulation environment is based on our WINTERsim system-level framework\footnote{http://winter-group.net/downloads/} and allows us implementing the above interference control mechanisms explicitly. We also model the airplane as a mobile object, which is simulated live with $25$ LTE cells for the time it takes the airplane to clear the network coverage area. During this entire time, the incumbent reports airplane position to the network controller, which, in turn, implements the interference mitigation policies on the BSs. The BSs employ the conventional LTE power control logic to enforce a particular policy on their UEs over the consecutive radio frames. The entire system is compliant with effective LTE specifications and has been calibrated against the reference 3GPP scenarios in our past publications~\cite{And14_2}.

\subsection{Performance results and their interpretation}

\begin{figure}[!ht]
\centering
\includegraphics[width=0.90\columnwidth]{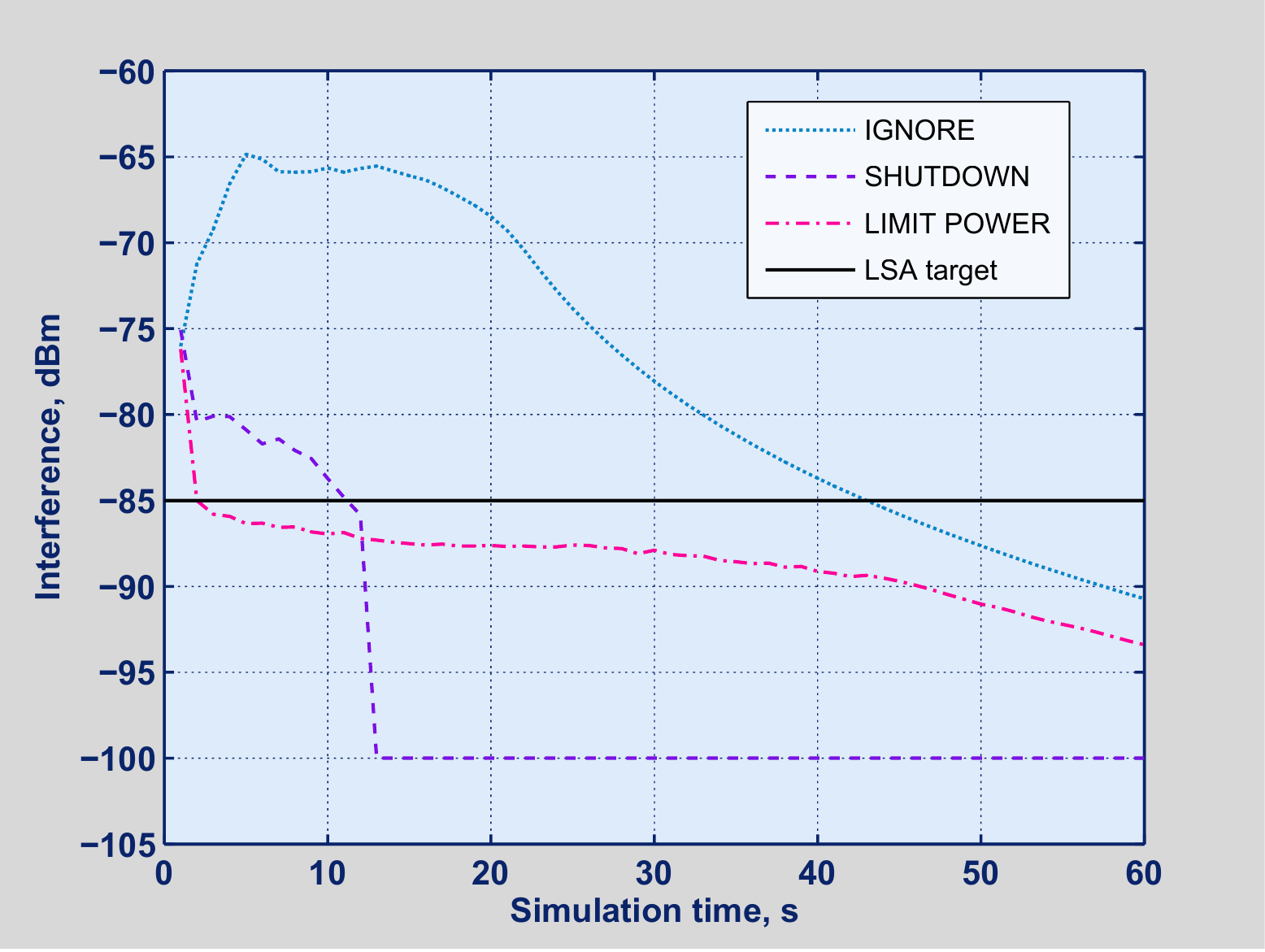}
\caption{Interference analysis of LSA operation}
\label{fig:interference}
\end{figure}

In what follows, our primary focus is on the performance analysis of highly dynamic LSA spectrum sharing and the respective measures to protect the reliable operation of the incumbent's systems. To this end, Fig.~\ref{fig:interference} demonstrates the levels of interference that the incumbent receives from the cellular users for all the considered control policies: IGNORE, SHUTDOWN, and LIMIT POWER. While IGNORE policy results in severe interference (as it does not reduce power), both SHUTDOWN and LIMIT POWER policies satisfy the interference requirements. It is important to note, however, that SHUTDOWN policy does not result in as prompt network reaction as one would expect. This is because the shutdown of a cell makes all of the associated UEs to join other nearby cells using close-to-maximum power, and as a result does not have the desired effect on the interference, even with excessive shutdown thresholds. By contrast, LIMIT POWER keeps the interference within limits as well as allows users to still transmit data, even with lower allocated power.

\begin{figure*}[!ht]
\begin{tabular}{ccc}
\includegraphics[width=0.66\columnwidth]{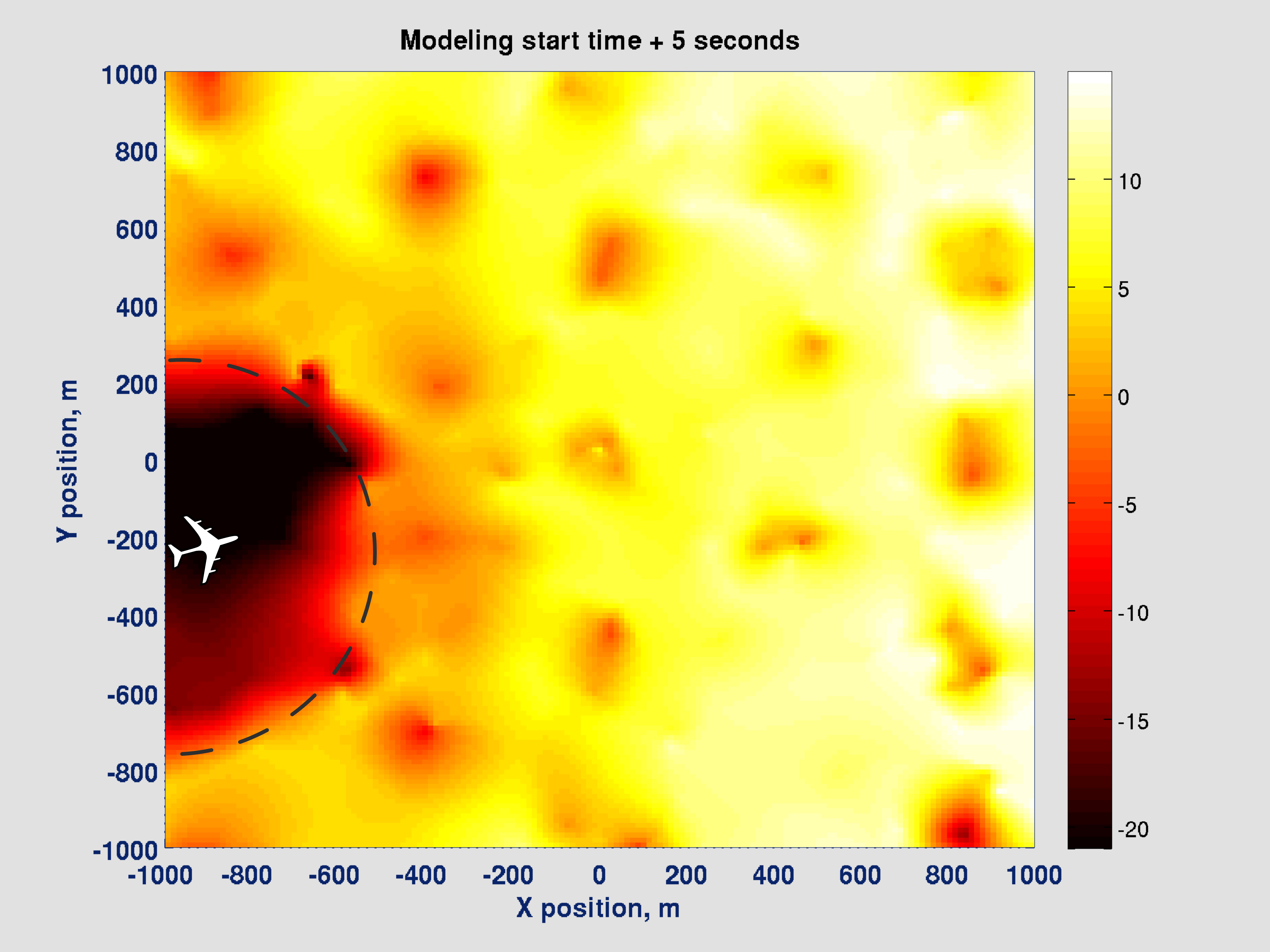} & \includegraphics[width=0.66\columnwidth]{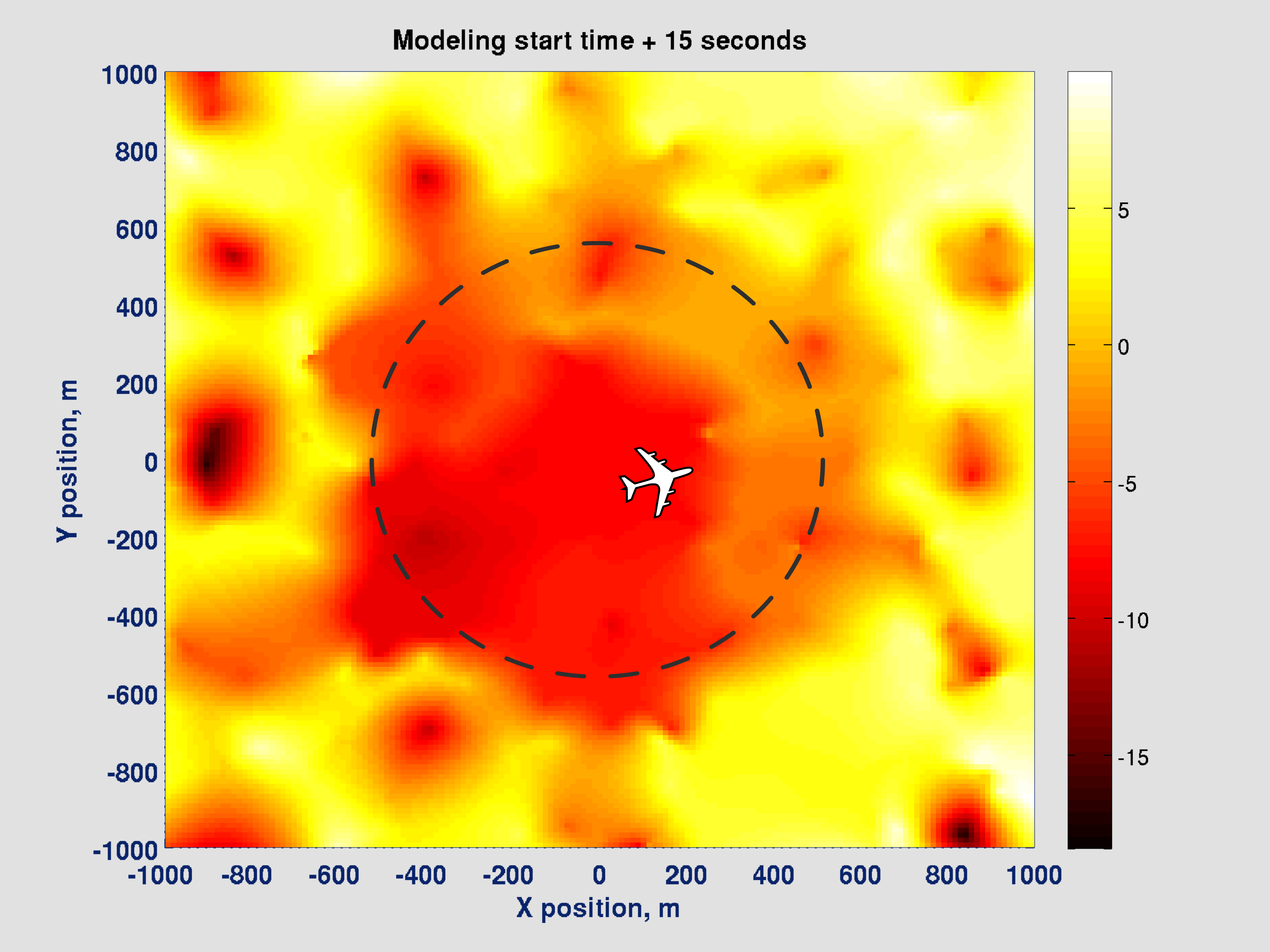}  & \includegraphics[width=0.66\columnwidth]{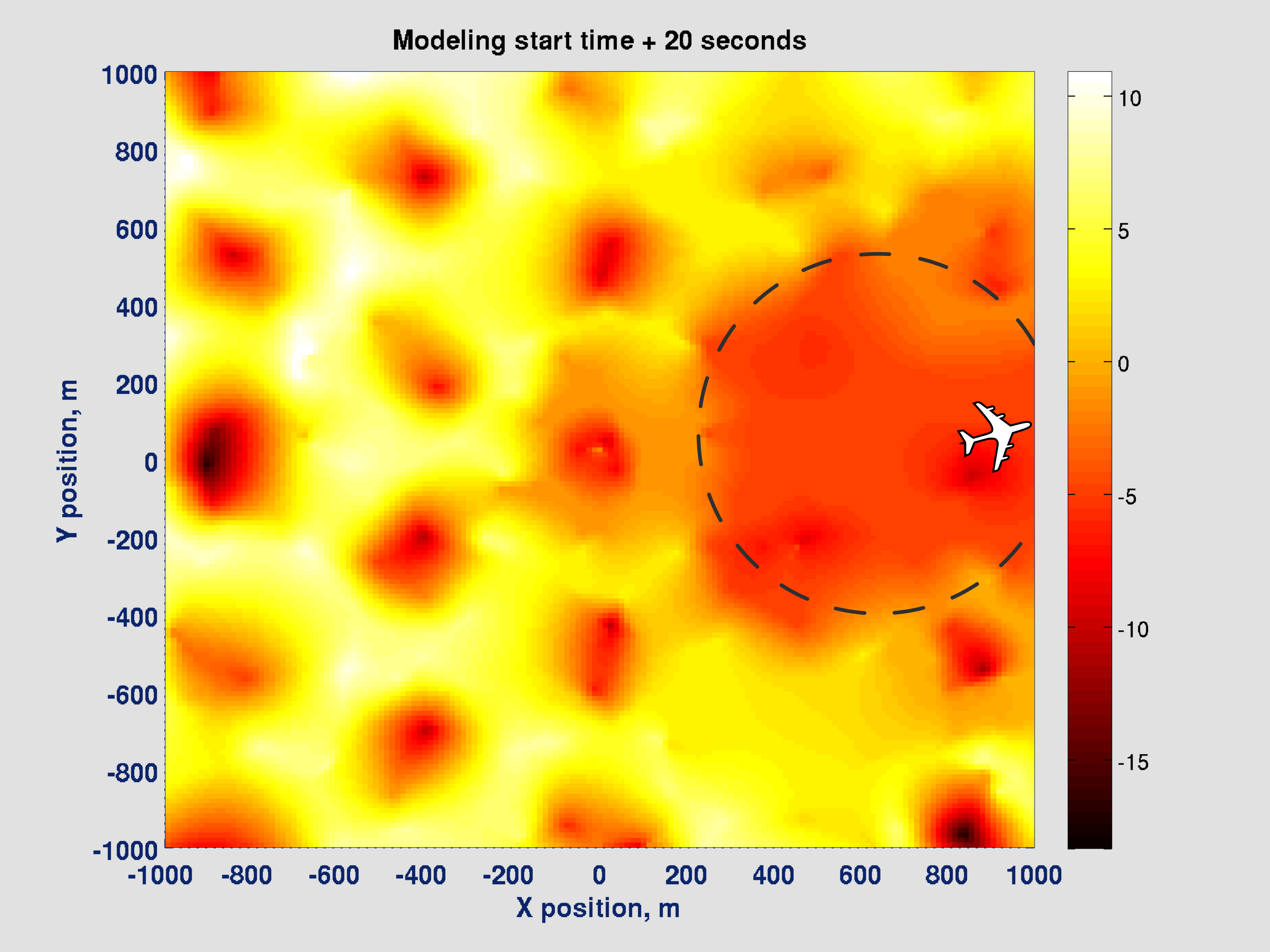}\tabularnewline
\end{tabular}
\caption{Performance evaluation: airplane moves across cellular network}
\label{fig:plane}
\end{figure*}

Further, we observe how the network responds to the airplane's mobility (see Fig.~\ref{fig:plane}). This is best visible for LIMIT POWER policy, as we can then monitor the UE's uplink power as the airplane moves across the network. On the heatmap plots, we clearly see that the airplane casts its radio "shadow", thus causing the surrounding users to decrease their transmit power. The approximate bounds of such shadow are shown with dashed lines and we see that it lags behind the airplane as it accelerates.

The snapshots in Fig.~\ref{fig:plane} are not regular in time since the airplane is gaining speed. What is crucial to note here, is that we never need to reduce the UE transmit power below -10 dBm to meet the interference constraints of the airplane. Hence, cell-center users can still continue utilizing the LSA bands as usual, sometimes even enjoying higher QoS (in the absence of cell-edge users). To make it happen, the network needs to employ a type of Proportional-Fair scheduler, that would allocate most of the resources to the cell-center users, since they are now the only ones with a reasonable SINR. In case of SHUTDOWN policy, the radio shadow would actually cause complete power-off of all the affected LSA cells and UEs, resulting in dramatic loss of capacity. Let us further investigate how much performance could be realistically gained by preferring LIMIT POWER over SHUTDOWN.

\begin{figure}[!ht]
\centering
\includegraphics[width=0.90\columnwidth]{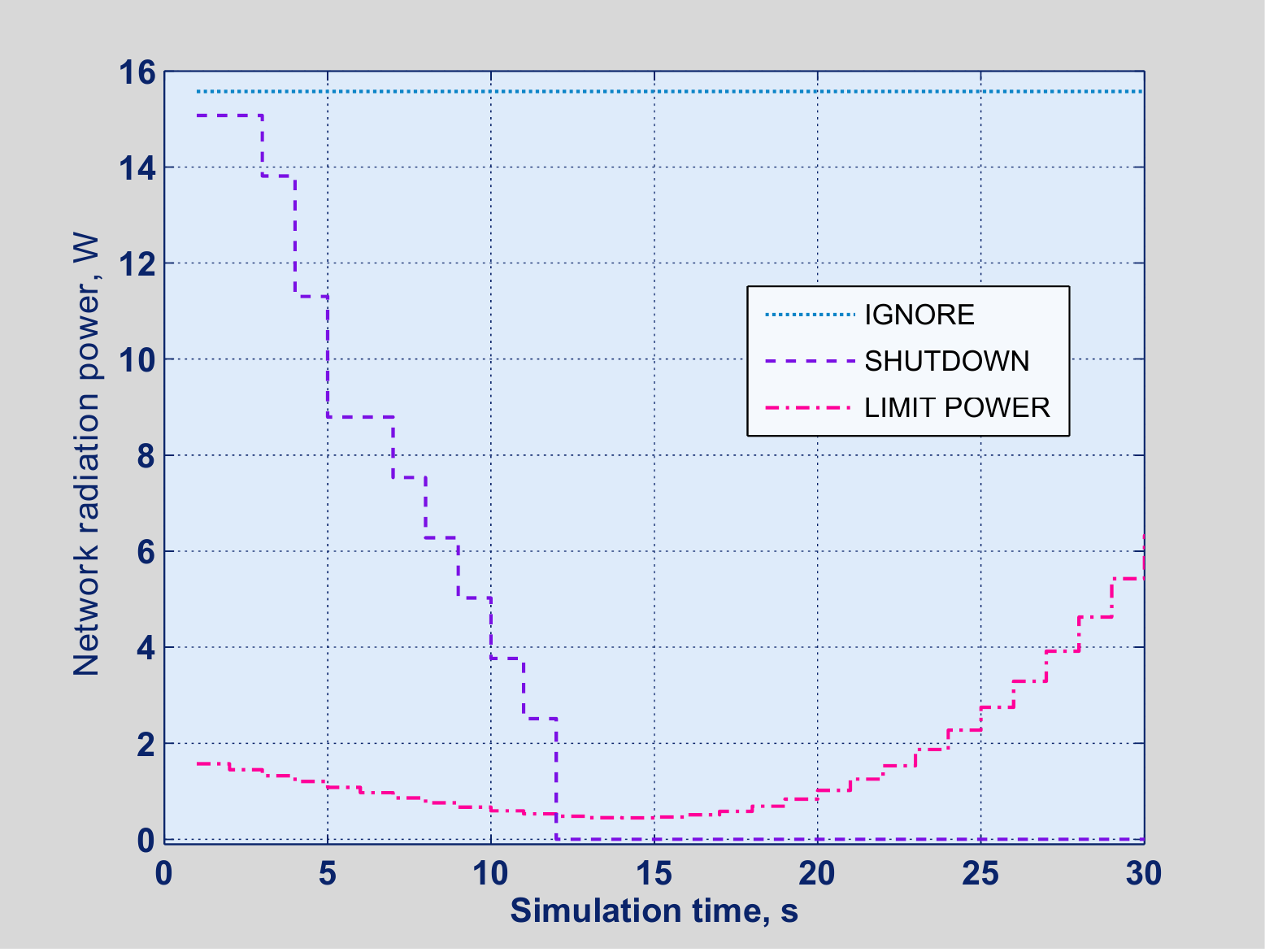}
\caption{Comparing alternative modes of LSA operation}
\label{fig:compare}
\end{figure}

One of the network's key performance indicators is how much energy has been radiated by the LSA cells altogether while the incumbent's airplane arrives/departs. This important metric can then be translated into throughput, subject to a particular scheduling policy, and Fig.~\ref{fig:compare} indicates how the three control policies compare against each other in this respect. While IGNORE policy is not practical, it gives a good idea of how much energy could have been radiated, should there be no airplanes. The SHUTDOWN and LIMIT POWER policies are practical, and their main difference is in the cost of added control interface complexity. For SHUTDOWN, it is a binary command, while LIMIT POWER requires regular updates of power thresholds on each BS.

In practical terms, LIMIT POWER, while not approaching IGNORE levels, still enables a considerable additional power that can be collectively radiated from the UEs in the immediate vicinity of the airport. As transmit power directly translates into throughput, more radiated energy generally allows for higher data rates. In case of LIMIT POWER policy, we observe that when the airplane is in the middle of the deployment, it affects most of the cells (see Fig.~\ref{fig:plane} to verify). Therefore, the radiated power is minimal at around $15$ seconds in Fig.~\ref{fig:compare}. 

It is also important to note that Fig.~\ref{fig:compare} reveals a step-wise behavior for LIMIT POWER. This is tightly connected with how the control mechanism operates. Essentially, the incumbent updates the reference position of the airplane every second in our scenario, and the cellular system responds by changing the power control settings accordingly. These changes are, in turn, reflecting on the interference levels as measured by the airplane and presented in Fig.~\ref{fig:interference}. In reality, the smoothness of the resulting steps will depend on the capacity/latency of the LSA control interface between the incumbent and the licensee.

\section{Current status and future evolution of LSA}

\subsection{Regulation and standardization update}

In Europe, LSA receives considerable political support from the European Commission (EC), which is the executive body of the European Union (EU). In particular, EC has asked its high-level advisory Radio Spectrum Policy Group (RSPG) to provide their opinion on LSA (RSPG 13-538). As the result, the EC has issued a standardization mandate on mobile/fixed communication networks (MFCN)~\cite{EC} in 2.3-2.4 GHz bands; the anticipated follow-up EC directives and decisions on LSA would be legally binding for the $28$ member states of EU. 

Currently, European adoption of voluntary spectrum sharing with LSA is facilitated by Conf\'{e}rence Europ\'{e}enne des Postes et des T\'{e}l\'{e}communications (CEPT), primarily for MFCN~\cite{Gun14}. The corresponding CEPT Working Group Frequency Management (WG FM) has established two project teams (named PT52 and PT53) to ensure the LSA framework is ready to be introduced to the market from a regulatory perspective. While the former addresses more specific implementation measures of LSA, the latter outlines its general aspects, including possible sharing arrangements and band-specific (if not dealt with by a specific project team) conditions for the implementation of the LSA that could be used as guidelines for CEPT administrations. 

Along these lines, PT52 has finalized their Decision (14)02 on harmonized technical and regulatory conditions for the use of 2.3-2.4 GHz bands for MFCN and delivered it to the European Communication Committee (ECC) of CEPT. Their other work included ECC Recommendation (14)04 on cross-border coordination for MFCN (and with other systems) in these bands. Presently, PT52 develops another ECC Recommendation to provide guidance to administrations in implementing a sharing framework between MFCN and PMSE (Programme Making and Special Events)~\cite{Rec1}, as well as prepares their response~\cite{Rec2} to the EC Mandate on 2.3-2.4 GHz bands. In turn, PT53 has delivered their ECC Report 205 on LSA that has been published in 2014.

From another end, European Telecommunications Standards Institute (ETSI), the main player in European standardization, has been tasked by other EC Mandate M/512 to enable the deployment and operation of CRS under LSA regime. Correspondingly, ETSI standardization has already issued technical specifications~\cite{TS113}, \cite{TS154} outlining system requirements and thus develops LSA system architecture~\cite{TS235}. This contributes to the overall picture produced by efficient interaction between ETSI standardization, CEPT regulation activities, and alignment with political objectives across Europe. Furthermore, several new LSA-related Work Item proposals have been submitted to 3GPP LTE Release 13 and it is currently under discussion how 3GPP will address these proposals.

Complementing European efforts behind licensed spectrum-sharing technology, similar approaches emerge in other geographical regions. In the US, the Federal Communications Commission (FCC) has issued their notice of proposed rulemaking (NPRM)~\cite{FCC} targeting the use of small cells in 3.5 GHz band based on a scheme, which is closely related to LSA. Accordingly, the spectrum is proposed to be managed by a spectrum access system (SAS) incorporating a dynamic database and, potentially, other interference mitigation techniques. While LSA foresees only two tiers of services (namely, incumbents and licensees), the FCC introduces the possibility for three tiers: incumbent access, priority access, and general authorized access (GAA).

In addition to regulation and standardization LSA activities, the respective research efforts are decisively getting momentum~\cite{Mat13}. For instance, the leading community conferences IEEE DYSPAN (Dynamic Spectrum Access Networks) and IEEE CROWNCOM (Cognitive Radio Oriented Wireless Networks) have recently reported an impressive number of research papers on LSA, ranging from spectrum occupancy measurements in 2.3-2.4 GHz bands to LSA trial demonstrations in live LTE network. This extensive research is supported by several visible project consortia, such as METIS-2, CORE+, CoMoRa, COHERENT, and some other 5G-PPP activities. All of the above creates fruitful soil for prompt development and adoption of LSA ecosystem, as well as subsequent efficient deployments.
\subsection{Summary on LSA performance promise}

While we expect additional global spectrum to be allocated for future mobile services as the result of the World Radiocommunication Conference 2015 (WRC-15), it is very likely that such new frequency bands will also encompass other legacy primary services. Here, LSA may help deliver considerable benefits to the MNOs when employed in conjunction with their primary allocation under exclusive licenses. However, to ensure that LSA will not enter in conflict with exclusive spectrum usage models, it should be based on effective market demand -- LSA should come as a complementary solution for accessing spectrum on particular bands, rather than a replacement to conventional exclusive access.

In the future, LSA might also add to the starting separation between the MNOs and the actual spectrum owners, with the latter offering more capacity to the former together with the associated spectrum availability guarantees. Correspondingly, extremely flexible and adaptable LSA implementations will be required, and our work may serve as an important building block to make it happen. Understanding that LSA may eventually alter the rules of competition, most MNOs have generally adopted the careful neutral behavior expecting when its potential benefits and associated limitations will become more clear. Seeking to resolve their concerns, this article sheds light on the expected LSA operation in a characteristic highly dynamic scenario. 



\bibliographystyle{ieeetr}
\bibliography{refs}

\section*{Authors' Biographies}

\textbf{Aleksei Ponomarenko-Timofeev} (aleksei.ponomarenko-timofeev@tut.fi) is a Researcher in the Department of Electronics and Communications Engineering at Tampere University of Technology, Finland. He received his B.Sc. degree from St. Petersburg State University of Aerospace Instrumentation, St. Petersburg, Russia. Aleksei co-authored a number of publications on spectrum access techniques, LSA framework, as well as H2H and M2M communications.

\textbf{Alexander Pyattaev} (alexander.pyattaev@tut.fi) is a Ph.D. Candidate in the Department of Electronics and Communications Engineering at Tampere University of Technology, Finland. He received his B.Sc. degree from St. Petersburg State University of Telecommunications, Russia, and his M.Sc. degree from Tampere University of Technology. Alexander has publications on a variety of networking-related topics in internationally recognized venues, as well as several technology patents. His primary research interest lies in the area of future wireless networks: shared spectrum access, smart RAT selection and flexible, adaptive topologies.

\textbf{Sergey Andreev} (sergey.andreev@tut.fi) is a Senior Research Scientist in the Department of Electronics and Communications Engineering at Tampere University of Technology, Finland. He received the Specialist degree (2006) and the Cand.Sc. degree (2009) both from St. Petersburg State University of Aerospace Instrumentation, St. Petersburg, Russia, as well as the Ph.D. degree (2012) from Tampere University of Technology. Sergey (co-)authored more than 90 published research works on wireless communications, energy efficiency, heterogeneous networking, cooperative communications, and machine-to-machine applications.

\textbf{Yevgeni Koucheryavy} (yk@cs.tut.fi) is a Full Professor and Lab Director at the Department of Electronics and Communications Engineering of Tampere University of Technology (TUT), Finland. He received his Ph.D. degree (2004) from TUT. He is the author of numerous publications in the field of advanced wired and wireless networking and communications. His current research interests include various aspects in heterogeneous wireless communication networks and systems, the Internet of Things and its standardization, as well as nanocommunications. He is Associate Technical Editor of IEEE Communications Magazine and Editor of IEEE Communications Surveys and Tutorials.

\textbf{Markus Mueck} (markus.dominik.mueck@intel.com) oversees Intel's technology development, standardization and partnerships in the field of spectrum sharing. In this capacity, he has contributed to standardization and regulatory efforts on various topics including spectrum sharing within numerous industry standards bodies, including ETSI, 3GPP, IEEE, the Wireless Innovation Forum and CEPT. Dr. Mueck is an adjunct professor of engineering at Macquarie University, Sydney, Australia, he acts as ETSI Board Member supported by INTEL and as general Chairman of ETSI RRS Technical Body (Software Radio and Cognitive Radio Standardization). He has earned engineering degrees from the University of Stuttgart, Germany and the Ecole Nationale Sup\'{e}rieure des T\'{e}l\'{e}communications (ENST) in Paris, France, as well as a doctorate degree of ENST in Communications.

\textbf{Ingolf Karls} (ingolf.karls@intel.com), Intel Deutschland GmbH, is program manager in Next Generation and Standards (NGS) team in Intel's Communication and Devices Group (iCDG). He has accomplished many projects in mobile communication at Siemens AG, Infineon Technologies AG and Intel for more than 20 years. He managed the operation of 14 national and European public funded research and development projects in European FP7. He contributed to national and international regulation and standardization bodies like 3GPP, BITKOM, DLNA, ETSI, IEEE, OMA and OMTP. He holds several patents in the area of wireless communication. His current research and development interests are the evolution of LTE and new innovative air interface solutions for 5G wireless technologies and standards and their translation into products.


\end{document}